\documentclass[aps,onecolumn,
longbibliography, 
notitlepage]{revtex4-2}
\usepackage[T1]{fontenc}
\usepackage[utf8]{inputenc}
\usepackage{color, bm}
\usepackage{amsmath}
\usepackage{graphicx}
\usepackage{subcaption}
\usepackage{tikz}
\usetikzlibrary{shapes,arrows,positioning,shadows}
\usetikzlibrary{arrows.meta}
\makeatletter

\usepackage[normalem]{ulem}
\usepackage[colorlinks=true,linkcolor=MidnightBlue,urlcolor=black,citecolor=MidnightBlue,anchorcolor=MidnightBlue]{hyperref}\usepackage[dvipsnames]{xcolor}

\usepackage{chngcntr}
\usepackage{multirow}
\makeatother
\usepackage{babel}
\begin{document}
\title{Optimal transport \textcolor{black}{and control} of an active particle near a plane wall}
\author{Utkarsh Maurya}
\affiliation{Department of Physics, Indian Institute of Technology Madras, Chennai, India}
\author{Kavya Swaminathan}
\affiliation{Department of Physics, Indian Institute of Technology Madras, Chennai, India}
\author{Ejaz Ashraf}
\affiliation{Department of Physics, Indian Institute of Technology Madras, Chennai, India}
\author{Rajesh Singh}
\affiliation{Department of Physics, Indian Institute of Technology Madras, Chennai, India}

\begin{abstract}
The control of active colloidal particles 
via optical traps is a cornerstone 
for
research of matter at the micron and nanometer scale. 
A central challenge in this domain is the derivation of optimal transport protocols that minimize the mean work required to move a particle over a finite-time interval. 
Here,
we present the Ritz method in which open-loop protocols 
are constructed from a global basis of Chebyshev polynomials. 
\textcolor{black}{The protocols are optimized using either a genetic algorithm or a gradient-based method.} 
We apply the method to study optimal transport of an active particle, which is
modeled as a
force-dipole (or a stresslet) near a no-slip wall.
The methodology is validated in the limits of zero activity and infinite wall separation, where it successfully recovers the known analytical protocols and 
 the theoretical minimum work.
Crucially, we demonstrate that the presence of the activity 
breaks the time-reversal symmetry of the optimal protocol found. 
This symmetry breaking is shown to be a complex function of the transport direction and the particle's intrinsic activity. Because the presented approach requires only the capability to simulate stochastic trajectories, it offers a robust, principled framework for optimizing transport protocols in complex fluid environments that remain inaccessible to exact analytical treatment.
\end{abstract}

\maketitle

\section{Introduction}
\label{sec:introduction}
The minimization of thermodynamic work during the finite-time 
transport of colloidal particles is a canonical problem in 
stochastic thermodynamics \cite{peliti2021stochastic, siraishi2023, seifert_2025}.
Optical tweezers provide a natural means of confining and transporting such
particles~\cite{ashkin2018nobel}, and the question of how to do so efficiently is both
practically relevant and theoretically rich.
In a bulk fluid environment, the analytical solution 
was derived in a seminal paper by Schmiedl and Schmiedl--Seifert \cite{schmiedl2007optimal}, which established that the optimal transport protocol for a passive Brownian particle in a translating harmonic trap is a linear ramp bounded by jump discontinuities. 
Active colloidal particles, which dissipate
energy from their environment to create mechanical disturbance around them, have attracted sustained interest as model systems for nonequilibrium
physics~\cite{Ramaswamy2010,Marchetti2013, vrugt2025exactly}.
Deriving protocols for 
the transport of active colloidal particles
is topic of recent investigations using theory and experiments \cite{garcia2025optimal, loos2024universal, monter2025optimal, zhong2024, davis2024active, cocconi2023, casert2024learning, gupta2023, shankar2022optimal, welker2026accuracy, olsen2026self, baldovinPRE2026}. 
Current studies 
focus on particles in bulk fluid, while in many practical and experimental settings
particles are frequently confined and transported in the vicinity of solid boundaries. 

Near a solid boundary, the thermodynamic work done in transporting active particles is modified by two distinct hydrodynamic phenomena that are absent in the bulk. First, the spatially dependent friction
suppresses the particle's diffusion as it approaches the wall.
Second, the motion of the active particle is modified due to hydrodynamic interactions with the no-slip boundary. When modeled as a stresslet (a force
dipole) ~\cite{batchelor1970stress, lauga2020fluid}, an active particle generates a flow field that
is modified near the wall via the Blake image system
~\cite{blake1971c}.
Crucially, the resulting wall-induced drift 
is highly
direction-dependent: extensile swimmers (pushers) are
hydrodynamically repelled from the boundary, whereas
contractile swimmers (pullers) are attracted.

In this paper, we demonstrate how the interplay of spatially dependent mobility and asymmetric active drift fundamentally modifies the stochastic thermodynamics of near-wall transport of active particles. We show that the boundary-induced effects profoundly distort the minimum-work optimal control protocols. 
To this end, we note that
the coupling between the spatially varying Brenner friction and the wall-induced active drift introduces severe nonlinearities into the governing equations, rendering exact analytical solutions for the minimum-work protocols intractable. 

A variety of numerical methods have been developed
for optimising time-dependent 
protocols in stochastic systems, including genetic algorithm \cite{maity2026emergent, back1996evolutionary, holland1992genetic,such2017deep, hartl2021microswimmers, whitelam2023demon, whitelam2025benchmark, montana1989training, mitchell1998introduction}, and gradient-based
algorithms~\cite{engel2023optimal, whitelam2025benchmark, davis2026optimal, rengifo2025}.
Genetic algorithms (GAs) are particularly well suited to problems of this kind: they
require only the ability to evaluate the objective function, are robust to noise, and
impose no smoothness requirements on the protocol ansatz \cite{whitelam2023demon, whitelam2025benchmark}.
\textcolor{black}{Gradient-based optimizers are faster than GAs and can manage noisy objective function with appropriate regularisation. At the same time, the two disparate methods can be used to compare results when no analytical solution for the protocol exists. Thus, in our method,
we allow the user to choose either GA or a gradient-based method (Adam \cite{kingma2014adam}) 
to find optimal paths. Unless specified otherwise, the default numerical method used in this paper is the gradient-based Adam optimiser \cite{kingma2014adam}.  }
Our numerical implementation is validated against the Schmiedl--Seifert exact solution \cite{schmiedl2007optimal} for a particle in the bulk. The numerical framework is then applied across
a systematic grid of initial 
wall separations $H_0$ and 
activity parameter $\alpha$, for both away-from-wall and towards-wall protocols.
\textcolor{black}{In addition, we find that the optimal 
protocols obtained using gradient-based optimiser and GA coincide.} \\

The rest of the paper is organised as follows.
First, in Section~\ref{sec:model}, we describe
the physical model, including the Brenner mobility and 
the stresslet active drift.
This section also describes the Ritz method, 
Chebyshev polynomials for protocol representation, 
and descriptions of two possible optimiser: (i) genetic algorithm and (ii)  gradient-based method.
Second, in Section~\ref{sec:results}, we 
presents our
results for open-loop protocols describing optimal transport of an active particle near a plane no-slip wall. 
\textcolor{black}{Here, we present protocols for both transport towards and away from the wall. Importantly, in this section, 
we show protocol distortion and broken time-reversal symmetry due to activity.} 
Finally, 
in section~\ref{sec:summary} we 
summarises the findings and suggest directions for future work.

\begin{figure}[t]
   \includegraphics[width=0.96\textwidth]{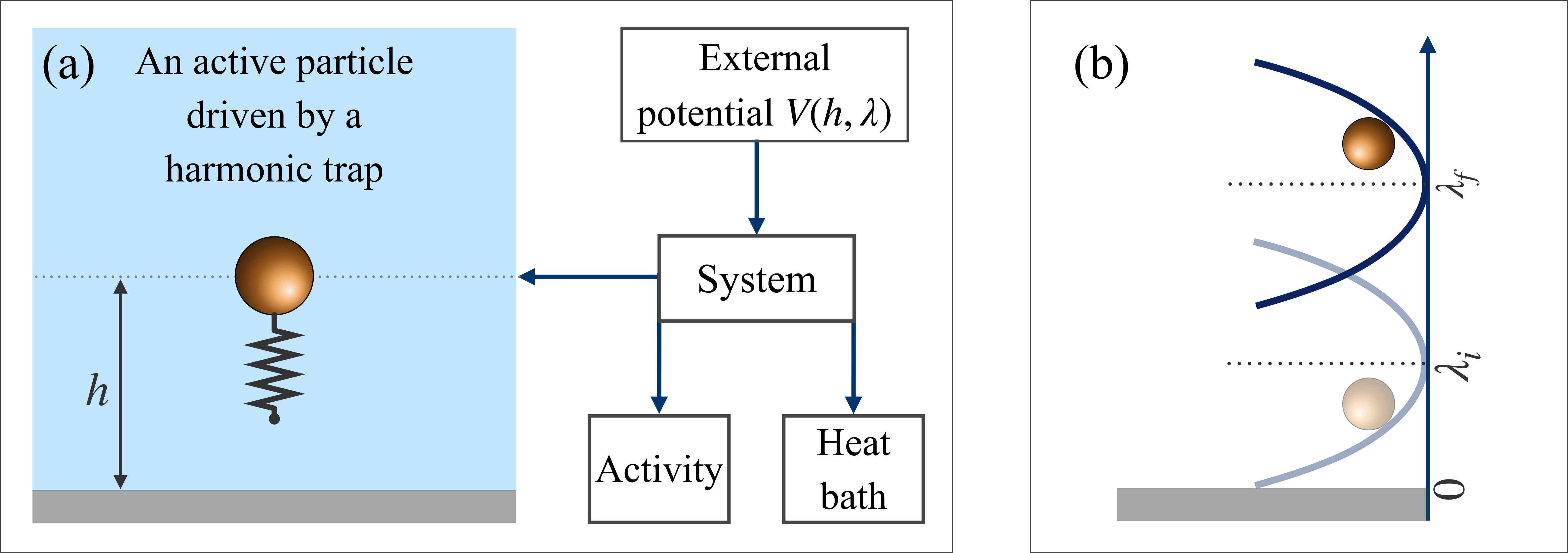}
  \caption{\textcolor{black}{A schematic diagram of the system}. (a) An active colloidal particle (colored sphere) is driven by a time-varying harmonic potential $V(h,\lambda)$.
  \textcolor{black}{The location of the particle is $h(t)$, 
  while the location of the center of the harmonic potential is $\lambda(t)$. In panel (b), $\lambda_i$ and $\lambda_f$ denote the center of the trap for
  the two terminals of the \emph{away} protocol, while 
  the plane wall is located at the origin.} 
  The particle exchanges energy with the trap, dissipates heat to the thermal
  bath, and self-generates drift through internal activity.
  }
  \label{fig:langevin_schematic}
\end{figure}

\section{Model and Methodology}
\label{sec:model}

\subsection{Model}

A spherical active colloidal particle of radius $b$ is immersed in a Newtonian
fluid of viscosity $\eta$ at temperature $T$ at height $h$ above an infinite
planar no-slip wall.
A schematic is given in Fig.~(\ref{fig:langevin_schematic}). 
The particle is confined in a harmonic optical trap of stiffness $k$ whose center
$\lambda(t)$ is the control parameter.
Explicitly, the potential is given as:
 \begin{align}
      V(h, \lambda) = \frac{k}{2} 
    \left[
    h - \lambda(t)
    \right]^2.
 \end{align}
The task is to move $\lambda$ from $\lambda_i = H_0 b$ to
$\lambda_f = (H_0 + \Delta H)b$ in a fixed time $t_f$ while
minimising the mean thermodynamic work.
We restrict attention to the wall-normal direction $h$, which captures the dominant
physics of wall-proximity effects \cite{lauga2020fluid}.
%
%
%
%
The dynamics of the active colloidal particle is given 
according to the overdamped Langevin equation \cite{gardiner1985handbook}:
\begin{align}
\label{eq:mainLE}
    \dot{h} &= -\mu(h)\,\partial_h V(h,\lambda ) + v^{\mathcal A}(h) + \sqrt{2D(h)}\,\xi(t)
    = A(h) +  \sqrt{2D(h)}\,\xi(t)
\end{align}
Here, $\mu(h)$ is the spatially dependent mobility and $D(h) = \mu(h)k_BT$ is the
local diffusion coefficient from the Einstein relation,
with $k_B$ as the Boltzmann constant~\cite{kubo1966fluctuation}.
Active velocity $v^{\mathcal{A}}(h)$ is defined below, while 
$A(h)$ is the total deterministic
drift. 
The stochastic variable $\xi(t)$ has zero mean: $\langle\xi(t)\rangle = 0$ and no temporal correlation: 
$\langle\xi(t)\xi(t')\rangle = \delta(t-t')$.
Thus, it is referred to as a `white' noise. 
We note that $D(h)$ inherits the spatial dependence of $\mu(h)$; both the drift
and the noise amplitude are therefore position-dependent, and this must be accounted
for in the numerical integrator (see section~\ref{sec:optimizer}).

The bulk Stokes mobility of a sphere of radius $b$ in a fluid of viscosity $\eta$
is $\mu_0 = 1/(6\pi\eta b)$.
Near a no-slip wall, the backflow generated by the wall image system reduces this
value. For a particle near a solid wall, the mobility $\mu(h)$ follows 
from the Brenner formula \cite{brenner1961slow}:
\begin{equation}
    \mu(h) = \mu_0 \left[ 1 - \frac{9}{8}\frac{b}{h} + \frac{1}{8}\left(\frac{b}{h}\right)^3 \right]
    \label{eq:brenner}
\end{equation}
where $b$ is the particle radius. Proximity to the wall induces a deterministic active velocity \cite{turk2024fluctuating}:
\begin{equation}
    v^{\mathcal A}(h) = -\frac{s_0}{8} \left[ \left(\frac{b}{h}\right)^2 - \left(\frac{b}{h}\right)^4 \right]
\end{equation}
For a puller ($s_0 > 0$) the drift is directed towards the wall; for a pusher
($s_0 < 0$) it is directed away from the wall.
Both the Brenner correction and the active drift vanish as
$b/h \to 0$, recovering homogeneous bulk dynamics far from the wall.
\textcolor{black}{The choice of $v^A$ is motivated from experiments on active particles \cite{drescher2011}. 
This is a distinct choice of 
for the model of an active particle as compared to 
an AOUP (active Ornstein-Uhlenbeck particle) usually studied in literature
\cite{garcia2025optimal, kumari2020stochastic, seifert_2025, davis2024active}.
Our model accounts to
a class of experiments where activity can arise from interactions of an active particle
with a boundary. }\textcolor{black}{Throughout the paper, for simplicity, we choose: $b=1,\mu_0=1,k=1, k_BT=1$. 
}


\subsection{Ritz Method for Protocol Representation}

In this section, we present the protocol representation for the Ritz
method~\cite{gelfand2000calculus, kikuchi2020}. 
The goal is to find a protocol which minimizes the mean thermodynamic work.
The mean thermodynamic work $\langle W \rangle$ performed during a time interval $[0, t_f]$ is defined as \cite{sekimoto1998langevin, peliti2021stochastic, siraishi2023, seifert_2025}:
\begin{align}
    \langle W \rangle &= \int_0^{t_f} dt\, 
    \dot{\lambda}(t) \left\langle \frac{\partial V(h, \lambda(t))}{\partial \lambda} \right\rangle
    = -k \int_0^{t_f}\, dt \dot{\lambda}(t) 
    \left[\langle h(t) \rangle - \lambda(t)
    \right]
\end{align}
In complex scenarios involving surface interactions or persistence, an analytical solution for the optimal protocol $\lambda^*(t)$ is generally unavailable; the numerical approach described in Section~\ref{sec:optimizer} is therefore required.

The open-loop protocol is
written as an expansion in Chebyshev polynomials $T_n$~\cite{boyd2001chebyshev}:
\begin{equation}
    \lambda(t) = \sum_{n=0}^{N-1} a_n\, T_n\!\left(\frac{2t}{t_f} - 1\right).
    \label{eq:ritz}
\end{equation}
The boundary conditions $\lambda(0) = \lambda_i$ and $\lambda(t_f) = \lambda_f$
are enforced exactly; the numerical algorithm optimises the
coefficients $\{a_n\}$ to minimise $\langle W \rangle$. 
Chebyshev polynomials are a natural
basis for this problem: they impose no smoothness constraint at the endpoints, so
the jump discontinuities of Eq.~(\ref{eq:seifert_protocol}) emerge naturally
from the optimization rather than being imposed \textit{a priori}. This global
spectral representation reduces the functional minimization to a multivariate
optimization over a compact, discrete set of coefficients. In the
implementation, the boundary conditions are enforced by appending the fixed
endpoint values $\lambda_i$ and $\lambda_f$ directly to the protocol array of points
before every physics evaluation, rather than as constraints on the coefficients
themselves; the Chebyshev expansion thus parametrises strictly the interior of
the time domain ($0 < t < t_f$), and the jump discontinuities known to appear
at $t = 0^+$ and $t = t_f^-$ in optimal protocols emerge naturally.

The infinite series of Eq.\eqref{eq:ritz} is truncated at $N = 5$ basis functions. Extensive
preliminary sweeps over the range $N \in \{2, \dots, 10\}$ revealed that the
mean thermodynamic work is almost same in this domain within the margin of error, as shown in Fig. \ref{fig:hero_convergence2} We selected $N = 5$ because it
optimally minimizes the work relative to analytical bulk solutions; expanding
the basis size further strains the fixed computational budget of the genetic
algorithm or gradient descent without yielding additional thermodynamic savings. Thus, the optimizer algorithms operate entirely within a 5-dimensional search space, optimizing the
finite set of coefficients $\{a_n\}$ to minimize $\langle W\rangle$.

\subsection{Dimensionless formulation and parameter space}
\label{sec:dimless}



In this section, we define a set of dimensionless parameters which can be varied to study the system defined above. First, we define a dimensionless activity parameter $\alpha$ as:
\begin{equation}
 \qquad \alpha = \frac{s_0}{\mu_0\,k\, b}=\frac{s_0\,\tau}{b},\qquad \quad {\tau}= \frac{6\pi\eta b}{k} = \frac{1}{\mu_0 k}.
    \label{eq:dimless}
\end{equation}
Here, ${\tau}$ is the natural time-scale of the system, while $s_0$ is the scale of the activity. Thus, the parameter $\alpha$ is a dimensionless quantity that controls the activity of the particle. In the limit of $\alpha=0$, the particle is passive, while it is pusher for $\alpha<0$ and puller for $\alpha>0$.

We have non-dimensionalised all length by the particle radius $b$, such that:
\begin{align}
      H = \frac{h}{b},  \qquad\qquad
      \Lambda = \frac{\lambda}{b}.
\end{align}
Here, $H$ is the dimensionless height of the particle from the wall, while $\Lambda$ is the dimensionless location of the trap from the wall. As described above, $h=H_0b$ is the initial location of the particle. We need to move it by a distance of $\Delta Hb$. 
Of the several parameters that appear, any two may be chosen independently; it is
conventional to choose $H_0$ and $\alpha$ to characterize the wall distance and
the activity respectively.
The parameter grid is $H_0 \in \{2, 3, 10, 1000\}$ and
$\alpha \in \{-25, 0, 25\}$, giving 15 combinations.
$H_0 = 1000$ is the effective bulk limit; $\alpha = 0$ corresponds to a passive
Brownian particle; $\alpha < 0$ to a pusher (wall-repelled); $\alpha > 0$ to a puller
(wall-attracted).
The transport displacement is $\Delta H = 5$ and the protocol duration
$t_f = 2.0\tau $ throughout.

\subsection{Optimizing the protocol defined using Ritz method }
\label{sec:optimizer}

\textcolor{black}{In this section, we explain the two distinct methods used to find the optimal protocol.
Since an analytical solution for the optimal protocol governing the transport of an active particle in a complex environment is not feasible, it is essential to have complementary numerical methods to evaluate the validity of the protocol. To achieve this, we have implemented two distinct optimization techniques: (i) a gradient-based optimizer and (ii) a genetic algorithm. A brief description of both methods is provided below. Additionally, a flowchart illustrating the optimization process for both methods—(a) the genetic algorithm and (b) the Adam (gradient-based optimizer)—is presented in Fig. \ref{fig:flowchart}.
}

\textcolor{black}{
 Each protocol is evaluated by running $N_\text{traj}$ independent stochastic
trajectories and computing $\langle W\rangle$ as the fitness to be minimised.
Trajectories are integrated with Itô scheme \cite{lau2007state}:
\begin{align}
\label{eq:em}
    h^{(n+1)} &= h^{n} + \Delta t \left[ A(h^{(n)}) + \frac{\partial D(h^{(n)})}{\partial h} \right] + \sqrt{2 D(h^{(n)})} \,dB^{n}
\end{align}
where $A(h)$ is the total deterministic
drift, and $dB$ is 
the Wiener increment \cite{kloeden1992higher}. 
Note that an additional drift (proportional to the divergence of the diffusion coefficient) must be added to the update equation so that the distribution function reaches the Boltzmann distribution at long times in absence of activity \cite{lau2007state}.
The timestep is $\Delta t = 0.002$ with $N_\text{steps} = 1000$ steps per
trajectory over $t_f = 2.0$.
}


For GA, a population of $P = 150$ individuals is initialised from a Gaussian distribution,
with the zeroth Chebyshev coefficient biased toward the midpoint
$H_0 + \Delta H/2$ to seed plausible initial protocols.
Each individual encodes a full protocol via basis evaluation at each timestep. 
The top $10\%$ of individuals by work value are retained unchanged as elites.
The remaining offspring are produced by tournament selection with tournament
size $T_s = 10$, followed by Gaussian mutation with a generation-dependent scale
    $
    \sigma_\text{mut}(g) = 0.2 - 0.195\,\frac{g}{G-1},
    $
where $g$ is the current generation and $G$ the total number of generations in
the stage.
The scale decreases from $0.2$ to $0.005$, allowing broad exploration early and
fine-grained refinement near the end of each stage.

To balance computational cost against statistical accuracy, a two-stage progressive
fidelity scheme is used.
In Stage~1 ($G_1 = 30$ generations) each individual is evaluated at
$N_\text{traj} = 500$ trajectories.
The best individual from Stage~1 seeds Stage~2, in which the population is
re-initialised around this seed and evaluated at $N_\text{traj} = 2{,}000$
trajectories for $G_2 = 70$ generations.
Additionally, the top-$10\%$ elites in Stage~2 are re-evaluated at
$N_\text{traj} = 100{,}000$ trajectories before selection, providing
high-fidelity pressure near the optimum.
The total budget is $G = 100$ generations.
The optimization is run over the $4\times3$ parameter grid of
Section~\ref{sec:dimless} with $30$ independent trials per grid point.
\begin{figure}[t] 
    \resizebox{1\textwidth}{!}{%
    \begin{tikzpicture}[
        scale=0.5,
        node distance=0.9cm,
        block/.style={rectangle, draw, text width=12em, text centered,
                      rounded corners, minimum height=4em, font=\bfseries, fill=white},
        repeatblock/.style={rectangle, draw, text width=6.4em, text centered,
                            rounded corners=12pt, font=\bfseries, fill=white,
                            inner sep=8pt},
        line/.style={draw, -{Latex}, thick, color=black}
    ]
    
    \node [block, text width=18em] (param)
        {Protocol Parameterisation \\
        \normalfont\small Chebyshev coefficients $\{a_n\}$,
        $n=0,\dots,4$};
    
    \node [block, below left=1.8cm and -2.0cm of param, xshift=-4cm] (ga_init)
        {GA: Initialise \\
        \normalfont\small Random population (150 individuals)};
    
    \node [block, below=of ga_init] (ga_eval)
        {Evaluate \\ \normalfont\small
            Simulate trajectories (stage 1: 500, stage 2: 2000);
            obtain $\langle W \rangle$};
    
    \node [block, below=of ga_eval] (ga_select)
        {Select \\ \normalfont\small
            Retain elites (10\%) and tournament selection};
    
    \node [block, below=of ga_select] (ga_mutate)
        {Mutate \\ \normalfont\small
            Add annealed Gaussian noise to coefficients};
    
    \node [repeatblock, right=1.0cm of ga_select, yshift=1.2cm] (ga_repeat)
        {Repeat \\ \normalfont\small
            $G_1=30$, $G_2=70$ generations;
            return best protocol};
    
    \path [line] (ga_init) -- (ga_eval);
    \path [line] (ga_eval) -- (ga_select);
    \path [line] (ga_select) -- (ga_mutate);
    \draw [line] (ga_mutate.south) -- ++(0,-0.8) -| (ga_repeat.south);
    \draw [line] (ga_repeat.north) |- ([yshift=1.2cm]ga_init.north) -- (ga_init.north);
    
    \node [block, below right=1.8cm and -2.0cm of param, xshift=4cm] (adam_init)
        {Adam: Initialise \\
        \normalfont\small Single set of coefficients (linear ramp)};
    
    \node [block, below=of adam_init] (adam_eval)
        {Evaluate \\ \normalfont\small
            Simulate a batch of trajectories (1000);
            compute $\langle W \rangle$ (loss)};
    
    \node [block, below=of adam_eval] (adam_grad)
        {Gradient \\ \normalfont\small
            Backpropagate through integrator
            ($\nabla_{\{a_n\}} \langle W \rangle$)};
    
    \node [block, below=of adam_grad] (adam_update)
        {Update \\ \normalfont\small
            Adam step with gradient clipping and
            cosine learning rate decay};
    
    \node [repeatblock, left=1.0cm of adam_grad, yshift=1.2cm] (adam_repeat)
        {Repeat \\ \normalfont\small
            $E=150$ epochs;
            return best protocol};
    
    \path [line] (adam_init) -- (adam_eval);
    \path [line] (adam_eval) -- (adam_grad);
    \path [line] (adam_grad) -- (adam_update);
    \draw [line] (adam_update.south) -- ++(0,-0.8) -| (adam_repeat.south);
    \draw [line] (adam_repeat.north) |- ([yshift=1.2cm]adam_init.north) -- (adam_init.north);
    
    \node [block, below=12cm of param, text width=26em, align=center] (final)
        {Final Evaluation \& Selection\\
        \normalfont\small
        Run both candidate protocols with 100,000 trajectories;
        compare $\langle W \rangle$ and protocol shapes;
        choose lower work (or verify agreement)};
    
    \coordinate (mid) at (0, -20.5);   
    
    \draw [line] (ga_repeat.east) -- ++(1.5,0) |- (mid);
    \draw [line] (adam_repeat.west) -- ++(-1.5,0) |- (mid);
    \draw [line] (mid) -- (final.north);
    
    \draw [line] (param.south) -- ++(0,-0.6) -| (ga_init.north);
    \draw [line] (param.south) -- ++(0,-0.6) -| (adam_init.north);
    
    \node [above=1.6cm of ga_init, font=\bfseries\large] {Genetic Algorithm};
    \node [above=1.6cm of adam_init, font=\bfseries\large] {Adam (gradient-based)};
    \end{tikzpicture}}
    \caption{\textcolor{black}{Flowchart for optimization of
    the protocol parametrized using Chebyshev coefficients in the Ritz method
    using two distinct methods: (a) genetic algorithm and (b) Adam (gradient-based).}
    In genetic algorithm, each generation evaluates
    $\langle W \rangle$ via stochastic trajectory simulation, retains the top
    elites, and produces the next generation by tournament selection and
    annealed Gaussian mutation. The loop runs for $G=100$ generations.}
    \label{fig:flowchart}
\end{figure}

\textcolor{black}{For our gradient-based method, we utilize the Adam optimizer \citep{kingma2014adam}. The steps involved in using Adam are outlined in Fig. \ref{fig:flowchart}. The protocol coefficients are initialized from a Gaussian distribution. The loss function $\langle W \rangle$ is evaluated over a batch of $N_\text{traj} = 1{,}000$ independent trajectories at each optimization step. Gradients are computed via backpropagation and used to update the protocol coefficients.}
\textcolor{black}{To ensure numerical stability in the presence of multiplicative noise, gradient clipping is applied to prevent divergent updates. The optimization is performed for $E = 150$ epochs, using cosine learning decay rate. The entire training procedure for all $30$ independent trials at a given grid point is vectorized, enabling concurrent optimization across trials. Unlike the GA, which maintains a population of candidate protocols, the Adam optimizer iteratively refines a single candidate protocol using gradient information. The optimization is run over the same $4\times3$ parameter grid as the GA.}

\textcolor{black}{At this point, it is worthwhile to compare the GA (genetic algorithm) with the gradient-based optimizer. Note that owing to the multiplicative nature of noise for the motion of an colloidal particle near the wall, the objective function is noisy. Thus, an out-of-the-box gradient-based method will not work for the current problem as the gradients will diverge. This necessitates regularisation by clipping the gradient to remain in certain bounds. No such regularisation is needed for GA.
}



\begin{figure*}[t]
    \includegraphics[width=\textwidth]{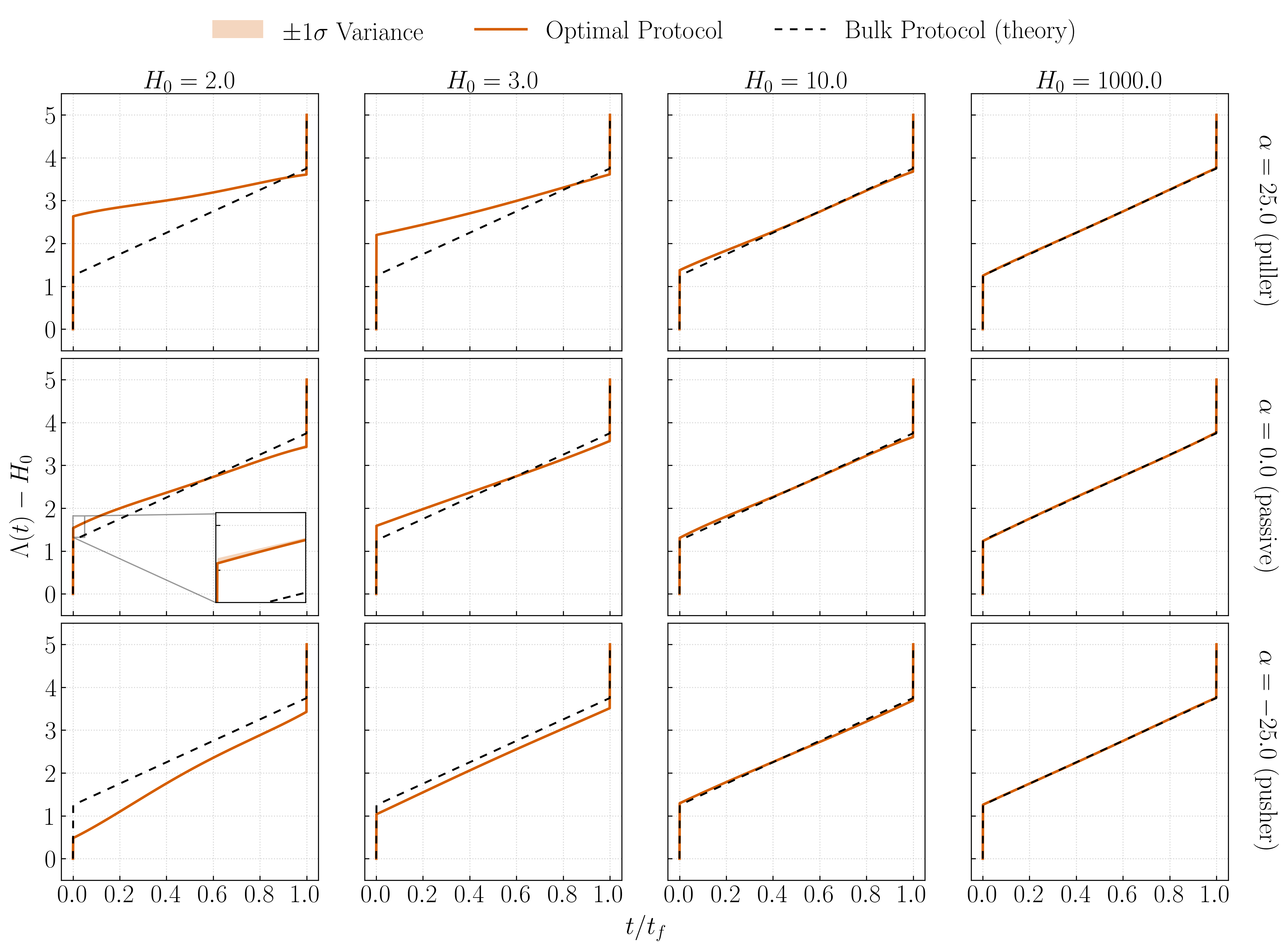}
    \caption{Optimal open-loop
    protocols for transport 
    \emph{away} from the wall, plotted as ${\Lambda}(t) - H_0$
versus normalised time $t/t_f$.
(1) Columns: initial distance from wall $H_0$; rows: activity $\alpha$.
(2) Solid orange: optimal protocol using Adam optimiser; shaded band: $\pm 1\sigma$ ensemble variance across 30 trials;
dashed black: Schmiedl--Seifert bulk prediction (Eq.~\ref{eq:seifert_protocol} of appendix \ref{app:seifert}).
(3) At $H_0 = 1000$ (rightmost column) the method recovers the bulk solution for all $\alpha$.
(4) Deviations from the bulk prediction grow as $H_0$ decreases and are modulated by activity.}
    \label{fig:protocol_grid}
\end{figure*}
\section{Optimal active transport near a wall}
\label{sec:results}
\subsection{Validation: recovery of the bulk limit}
\label{sec:validation}
The numerical 
method in section \ref{sec:model} is first validated against the analytical bulk limit (see appendix \ref{app:seifert} for the theoretical form of the bulk result \cite{schmiedl2007optimal}).
Fig.~\ref{fig:protocol_grid} shows the optimal protocols over the full $4\times3$
parameter grid for away-from-wall transport.
At $H_0 = 1000$ the optimized protocol (solid line) coincides with the
Schmiedl--Seifert prediction (dashed line) for all three values of $\alpha$: a linear ramp with
jump discontinuities at $t = 0$ and $t = t_f$.
The corresponding mean work agrees with $W^* = 6.25$ (Eq.~\ref{eq:seifert_work})
to within the statistical uncertainty of the retrospective evaluation, and the
$\alpha \neq 0$ cases recover the passive protocol since $v^{\mathcal A} \propto (b/h)^2 \to 0$
as $h \to \infty$.
This agreement in protocol shape, work value, and activity independence in the
far-field limit confirms the correctness of the numerical implementation.
Table~\ref{tab:work_values} gives $\langle W \rangle$ and $W_\text{SS}$
across the full parameter grid.
Here $\langle W \rangle$ is the mean work obtained using the Adam optimizer. 
$W_\text{SS}$ is the mean work obtained using the Schmiedl--Seifert protocol \cite{schmiedl2007optimal}. 
At $H_0 = 1000$ both quantities recover $W^* = 6.25$ and $\Delta W\%$
is consistent with zero for all $\alpha$. 
The magnitude of $\Delta W\%$ grows monotonically as the wall is approached and
is modulated by activity, as discussed in
Sections~\ref{sec:wallresults} and~\ref{sec:activity}.
The quantity
$\Delta W\%$ 
is defined in 
appendix \ref{app:seifert}. Ensemble Statistics for away-from-wall transport is given in appendix
\ref{sec:ensemeble}.

\begin{table}[t]
\caption{Minimum evolved work $\langle W \rangle $ (best of 30 trials),
work $W_\text{SS}$ obtained by running the bulk protocol
(see Eq.~\ref{eq:seifert_protocol}) through the near-wall simulator at
$100{,}000$ trajectories, and percentage difference $\Delta W\%$
(see Eq.~\ref{eq:deltaw}).
}
\label{tab:work_values}
\begin{ruledtabular}
\begin{tabular}{llcccc}
 & & \multicolumn{4}{c}{$H_0$} \\
\cline{3-6}
$\alpha$ & & $2.0$ & $3.0$  & $10.0$ & $1000.0$ \\
\hline
\multirow{3}{*}{$+25$ (puller)}
  & $\langle W \rangle/W^*$  & 1.466 & 1.324 & 1.064 & 1.000 \\
  & $W_\text{SS}/W^*$       & 1.534 & 1.363 & 1.065 & 1.000 \\
  & $\Delta W\%$             &  $-4.429$ &  $-2.831$ &  $-0.067$ &  $0.000$ \\
\hline
\multirow{3}{*}{$0$ (passive)}
  & $\langle W \rangle/W^*$  & 1.221 & 1.185 & 1.053 & 1.000 \\
  & $W_\text{SS}/W^*$       & 1.226 & 1.190 & 1.054 & 1.000 \\
  & $\Delta W\%$             &  $-0.446$ &  $-0.461$ &  $-0.049$ &  $-0.000$ \\
\hline
\multirow{3}{*}{$-25$ (pusher)}
  & $\langle W \rangle/W^*$  & 0.994 & 1.062 & 1.042 & 1.000 \\
  & $W_\text{SS}/W^*$       & 1.015 & 1.067 & 1.042 & 1.000 \\
  & $\Delta W\%$             &  $-2.102$ &  $-0.460$ &  $-0.037$ &  $0.000$ \\
\end{tabular}
\end{ruledtabular}
\end{table}
\begin{figure*}
    \centering
        \includegraphics[width=0.48\textwidth]{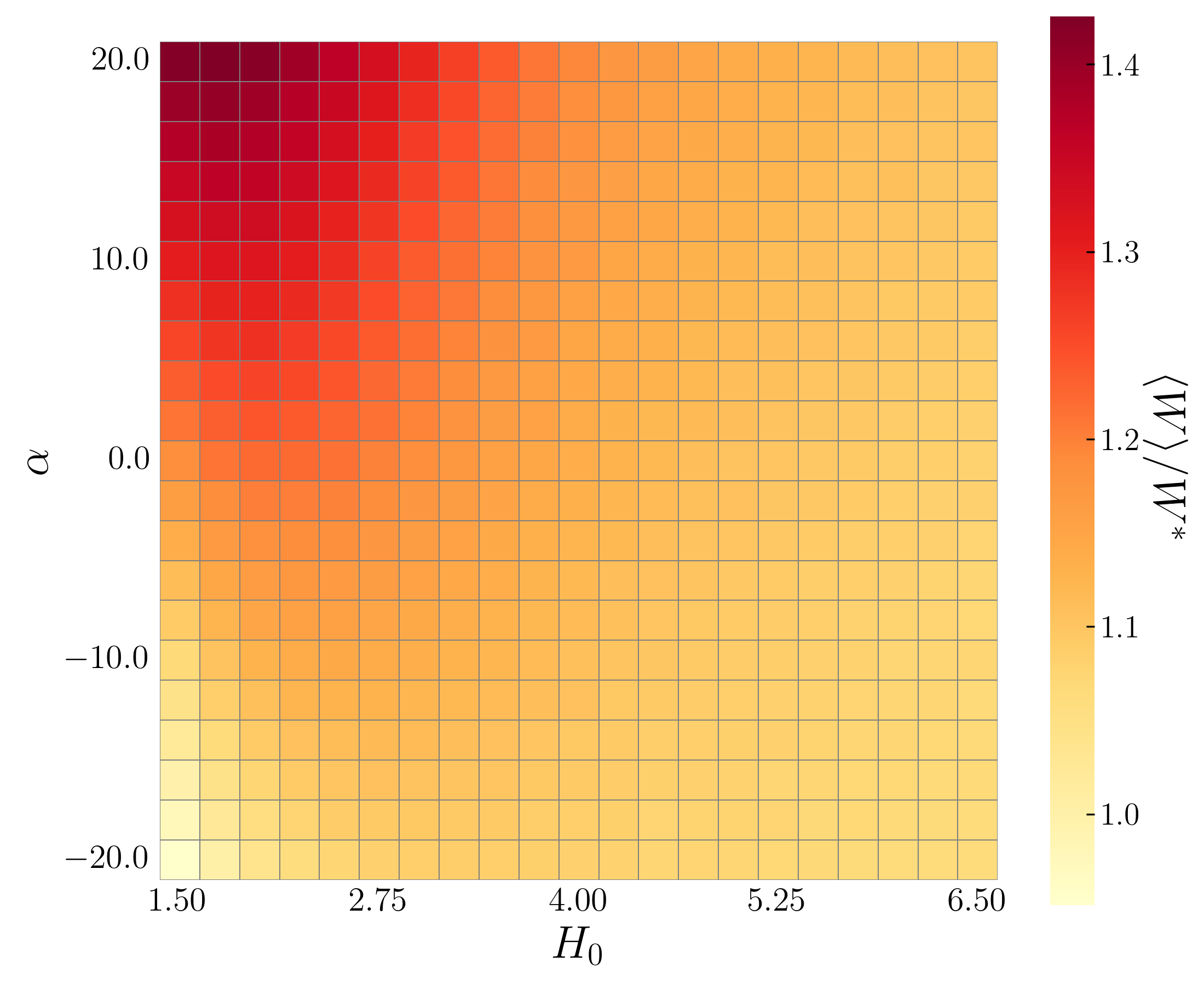}
    \hspace{0.9em}
        \includegraphics[width=0.48\textwidth]{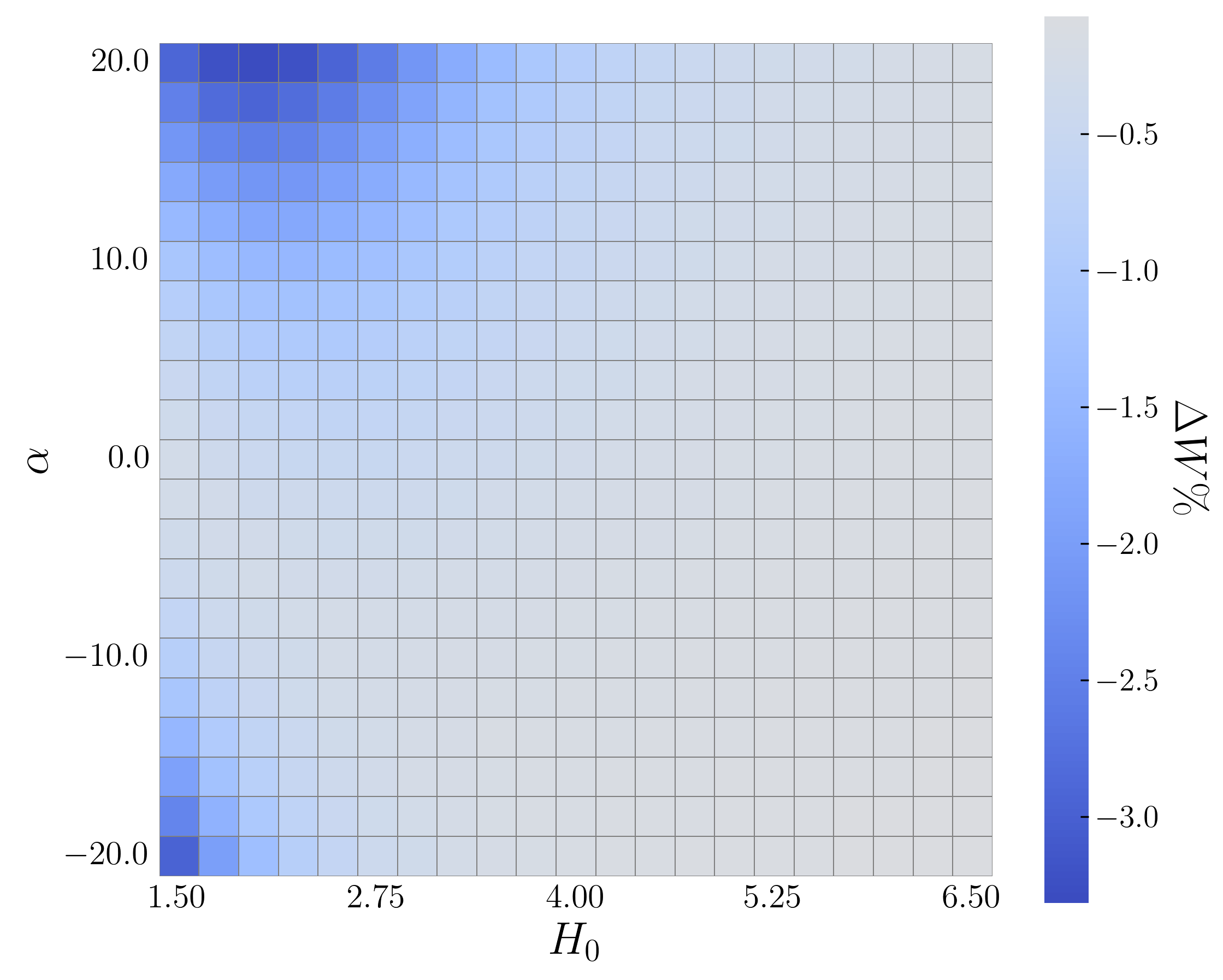}
    \caption{Heatmaps of ADAM optimization results for away-from-wall transport.
LEFT: mean thermodynamic work $\langle W \rangle$ as a function of
$H_0$ and $\alpha$. The work has been normalized with the bulk value $W^*$ defined in Eq.\eqref{eq:seifert_work}.
RIGHT: percentage difference $\Delta W\% = 100\times(\langle W \rangle_{} -
W_\text{SS})/W_\text{SS}$; negative values indicate that the ADAM
protocol costs less work than the Schmiedl--Seifert protocol under near-wall dynamics.
The magnitude of $\Delta W\%$ is largest at small $H_0$ where wall
effects are strongest.}
    \label{fig:heatmaps_away}
\end{figure*}
\subsection{Optimal protocols and work cost near the wall}
\label{sec:wallresults}
In this section, we study the optimal protocols for transport away from the wall.
We first examine the away protocol for the passive case $\alpha = 0$ (middle row of
Fig.~\ref{fig:protocol_grid}).
At $H_0 = 10$ the optimized protocol is nearly indistinguishable from the
bulk prediction. Deviations increase as $H_0$ decreases.
At $H_0 = 2$, where the Brenner correction, \textcolor{black}{
see Eq.\eqref{eq:brenner},
substantially reduces the
mobility of the particle,} the deviations are pronounced: the initial jump at $t = 0$ grows
substantially, the interior slope is much flatter than the linear bulk ramp,
and the protocol terminates with a steeper rise near $t = t_f$.

The physical origin of this \textcolor{black}{deviation from the bulk protocol can be understood as follows.
The bulk  
protocol, assuming uniform drag throughout, does not account for
this position-dependent mobility of the particle and is therefore suboptimal near the wall}.
Near the wall the particle lags further behind the trap center because the reduced
mobility limits its response to the applied force.
The optimal protocol compensates by front-loading a large initial jump 
and slowing the trap in the interior so that the
force-displacement product integrated over time, which determines
$\langle W \rangle$, is reduced. \textcolor{black}{Thus, the wall-induced corrections to mobility alters the protocol in non-trivial way.}

The pseudo-color-map of Fig.~\ref{fig:heatmaps_away} summarises these trends across the full parameter
space for $H_0 \in [1.5, 6.5]$.
Fig.~\ref{fig:heatmaps_away}a shows that $\langle W \rangle $ increases
monotonically as $H_0$ decreases, confirming that wall proximity always
incurs an energetic penalty.
Fig.~\ref{fig:heatmaps_away}b shows $\Delta W\%$ (Eq.~\ref{eq:deltaw}) for the
passive case; the magnitude grows as the wall is approached and falls to near zero
as $H_0 \to \infty$, consistent with the protocol shapes in
Fig.~\ref{fig:protocol_grid}.

\subsection{Effect of activity: pushers vs. pullers}
\label{sec:activity}
We now study protocols for the active ($\alpha\neq 0$) case in Fig.\eqref{fig:protocol_grid}. It is worth noting that activity and wall proximity enter the problem independently
through Eq.~(\ref{eq:mainLE}): the former via the $\alpha$-dependent drift term and
the latter via the Brenner mobility.
Their combined effect on $\langle W \rangle $ and $\Delta W\%$ is mapped
across the full $(H_0, \alpha)$ parameter space in Fig.~\ref{fig:heatmaps_away}.
For pullers ($\alpha > 0$, top row) activity, 
the stresslet drift is directed towards the wall, and thus a larger initial jump is needed
during away transport.
For the away journey, the particle spends its initial time in the
high activity region, 
so the trap must make an immediate large jump for puller activity to build
up a restoring force before the particle can respond. 
On the other hand, for a pusher ($\alpha<0$) the jump is smaller than the passive case. 
See Fig.~\ref{fig:protocol_grid} for away from wall transport protocols. 
This is because the activity of a pusher assists the away transport. 
Activity is thermodynamically beneficial when its direction coincides
with the transport direction. See Table~\ref{tab:work_values}.
\textcolor{black}{In Fig.~\ref{fig:protocol_grid} and the $\alpha$-axis of the heatmaps in
Fig.~\ref{fig:heatmaps_away}, we vary $\alpha$ at fixed $H_0$ for away transport of the particle.
Comparing rows of pushers and puller, the protocols differ from the passive case. 
}



\begin{figure*}[t]
    \centering
    \includegraphics[width=0.8\textwidth]{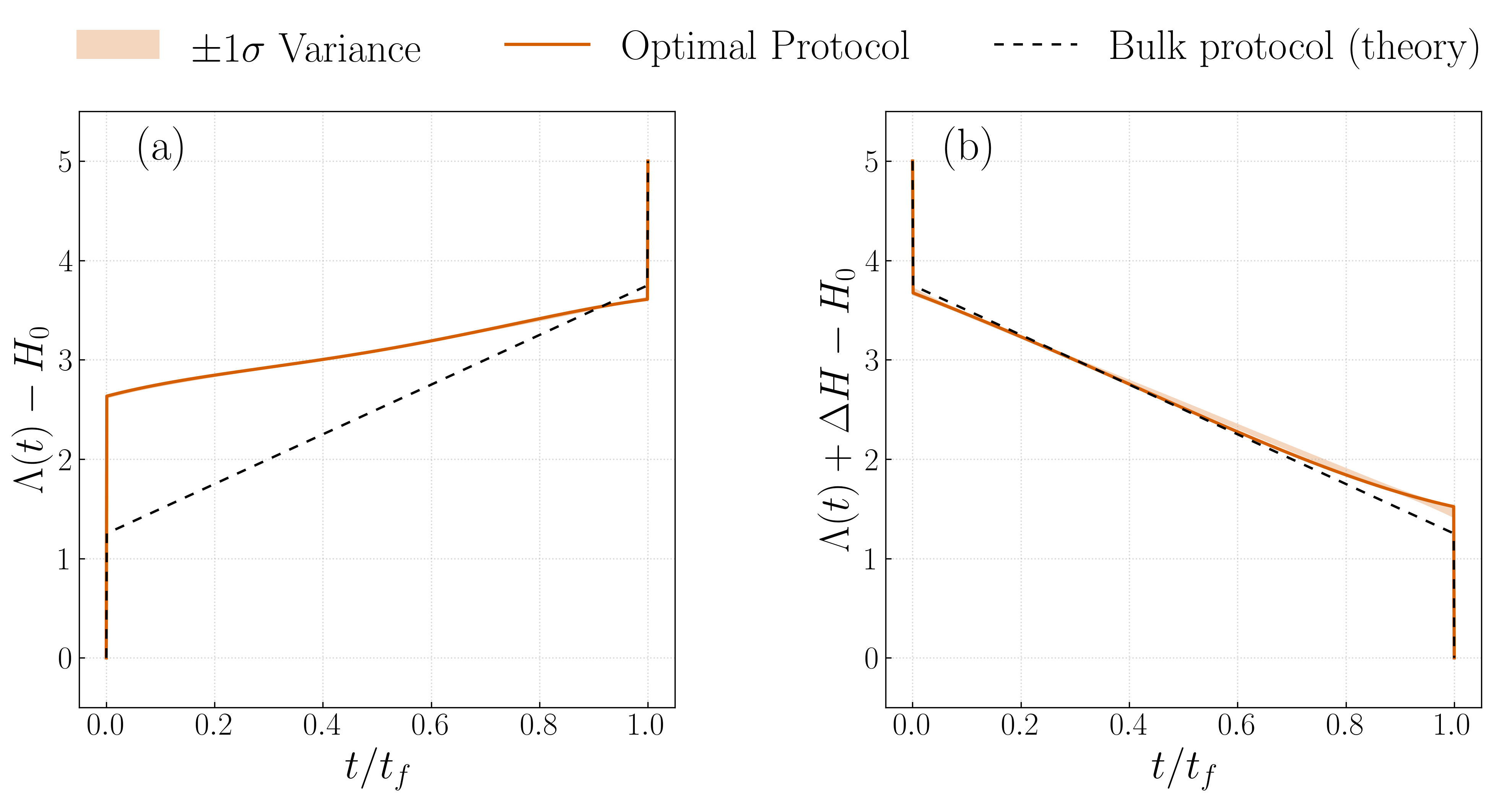}
    \caption{Symmetry breaking in the optimal protocol for a puller active particle
    ($\alpha = 25$).
    (a) Optimal away protocol with $H_0 = 2.0$. 
    (b) Optimal towards protocol with $H_0 = 7.0$.
    Solid orange: optimal protocol obtained in our numerical 
    method; shaded band: $\pm 1\sigma$ ensemble variance
    across 30 trials; dashed black: Schmiedl--Seifert bulk prediction.
    The away protocol, shown in Fig.~\ref{fig:symmetry}a, exhibits a large initial jump and a plateau-then-rise
    shape; while the towards protocol of Fig.~\ref{fig:symmetry}b tracks the bulk solution throughout most of the
    trajectory, deviating only near $t = t_f$.}
    \label{fig:symmetry}
\end{figure*}

\subsection{Symmetry breaking between open-loop protocols for away and towards transport}
\label{sec:symmetry}
In the bulk fluid, the 
open-loop 
optimal protocol possesses a time-reversal symmetry: the protocol
for transport from $H_0$ to $H_0 + \Delta H$ is the exact
time-reversal of the return protocol, and both achieve the same $W^*$
(Section~\ref{app:seifert} of the appendix).
This symmetry follows from the spatial uniformity of the bulk dynamics.
For the dynamics of an active particle near the wall,
this symmetry is broken.
When moving away, the particle begins in the high-drag and high-activity region and ends
in the near-bulk region and low activity.
The reverse happens for towards protocol.

Fig.~\ref{fig:symmetry} shows the optimal protocols for both directions at
$H_0 = 2$, $\alpha = 25$. The away protocol is shown in Fig.~\ref{fig:symmetry}a, while the towards protocol is shown in Fig.~\ref{fig:symmetry}b.
It can be seen that Fig.~\ref{fig:symmetry}a 
deviates strongly from the bulk prediction: there is a
pronounced jump at $t = 0$, to compensate for puller activity.
Fig.~\ref{fig:symmetry}b tracks the bulk prediction for most of the trajectory,
departing only near $t = t_f$ when the particle approaches the wall.
We note that the $\pm 1\sigma$ bands are narrow in both panels, so the asymmetry
is a systematic feature of the optimal solution rather than a statistical artefact.

For the away journey, the particle spends its initial time in the
suppressed-mobility region and high activity region due to puller activity. 
Thus, the trap must make an immediate large jump.
For the towards journey the particle starts where the Schmiedl--Seifert protocol is nearly
optimal, and corrections accumulate only in the final phase as the wall is
approached.
Not unexpectedly, applying the bulk Schmiedl--Seifert protocol for near-wall transport
incurs a much larger work penalty in the away direction than in the towards
direction. This direction-reversal of the activity effect, together with the protocol-shape
asymmetry, completes the characterization of the broken time-reversal symmetry.

\section{Summary and outlook}
\label{sec:summary}

In this work, we have presented a Ritz method
for computing minimum-work
optical trap protocols for an active colloidal particle near a no-slip wall.
The protocol is constructed using Chebyshev polynomials and it is  
\textcolor{black}{optimized using either a genetic algorithm
or a gradient-based method}. 
We apply the method to study optimal transport of an active particle,
which is modeled as  a stresslet \cite{lauga2020fluid} 
that generates a wall-induced drift, while
the mobility is suppressed by the
Brenner formula \cite{brenner1961slow}.
These two effects, absent from the Schmiedl--Seifert analysis, are the primary
sources of deviation from the bulk optimal protocol near the wall.

Three results stand out.
First, the methodology of section \ref{sec:model},
\textcolor{black}{which uses a 
gradient-based optimiser along with 
the Ritz parametrization of the protocol,}
recovers the Schmiedl--Seifert protocol and the analytical work value
$W^* = 6.25$ in the $H_0 \to \infty$, $\alpha = 0$ limit, validating the
numerical implementation.
Second, near-wall confinement distorts the optimal protocol and increases the mean
work; the method of section \ref{sec:model} yields $|\Delta W\%|$ of up to ${\sim}4.4\%$ relative to the Schmiedl--Seifert
protocol evaluated under near-wall dynamics at $H_0 = 2$, with the savings
growing as the particle approaches the wall.
Finally, the optimal protocols for away and towards transport break the time-reversal
symmetry in presence of activity: the away protocol requires large early corrections while the
towards protocol remains close to the bulk solution throughout most of the
trajectory, deviating only in the final phase.
The away and
towards costs, mapped across the full $(H_0, \alpha)$ parameter space, are given in
Figs.~\ref{fig:heatmaps_away} and~\ref{fig:heatmaps_towards} of the paper respectively.

Several simplifications in the present model merit comment.
The analysis is one-dimensional (wall-normal), and the lateral Brownian motion of
the particle has been neglected.
In addition, the orientation of the active particle has been considered to be fixed along the wall normal direction. 
\textcolor{black}{This effective one-dimensional motion is
an approximation which can arise due to large bottom-heaviness of active particles \cite{goldstein2015green}}. 
The wall correction used in this paper is a
leading-order far-field approximation; lubrication
corrections become important as $H \to 1$ and are not captured here.
The protocol is feedforward: no real-time measurement of the particle position is
used, and incorporating feedback would generally reduce dissipation
further~\cite{alvarado2026optimal, seifert_2025}.
The present study also treats a single particle; dynamics in
many-particle systems are not addressed, which suggests exciting direction for future work.

The Ritz method framework presented in this paper requires only the ability to simulate stochastic
trajectories and is therefore applicable to any
generic stochastic system. 
Natural extensions include three-dimensional transport with lateral fluctuations
and rotational--translational coupling, simultaneous optimization of trap stiffness
and position, and many-particle systems with hydrodynamic interactions.
The direction-dependent symmetry breaking reported here is a general feature of
systems with spatially inhomogeneous dynamics, where transport in opposite
directions samples the environment in reversed order.
\textcolor{black}{Application of the method to obtain minimum heat dissipated \cite{seifert_2025}  protocols 
and/or protocols which minimize the entropy production rate \cite{seifert_2025} 
suggest further avenues of investigations.}

\section*{Conflicts of interest}
There are no conflicts to declare.

 
\section*{Acknowledgments} 
RS thanks R. Adhikari, M. E. Cates, and
S. A. M. Loos for stimulating discussions.
We also thank anonymous referee for several suggestions which has led in improvement of the paper. 

\appendix

\section{Towards-wall protocols and heatmaps}
\label{app:towards}
\begin{figure*}[t]
    \centering
    \includegraphics[width=\textwidth]{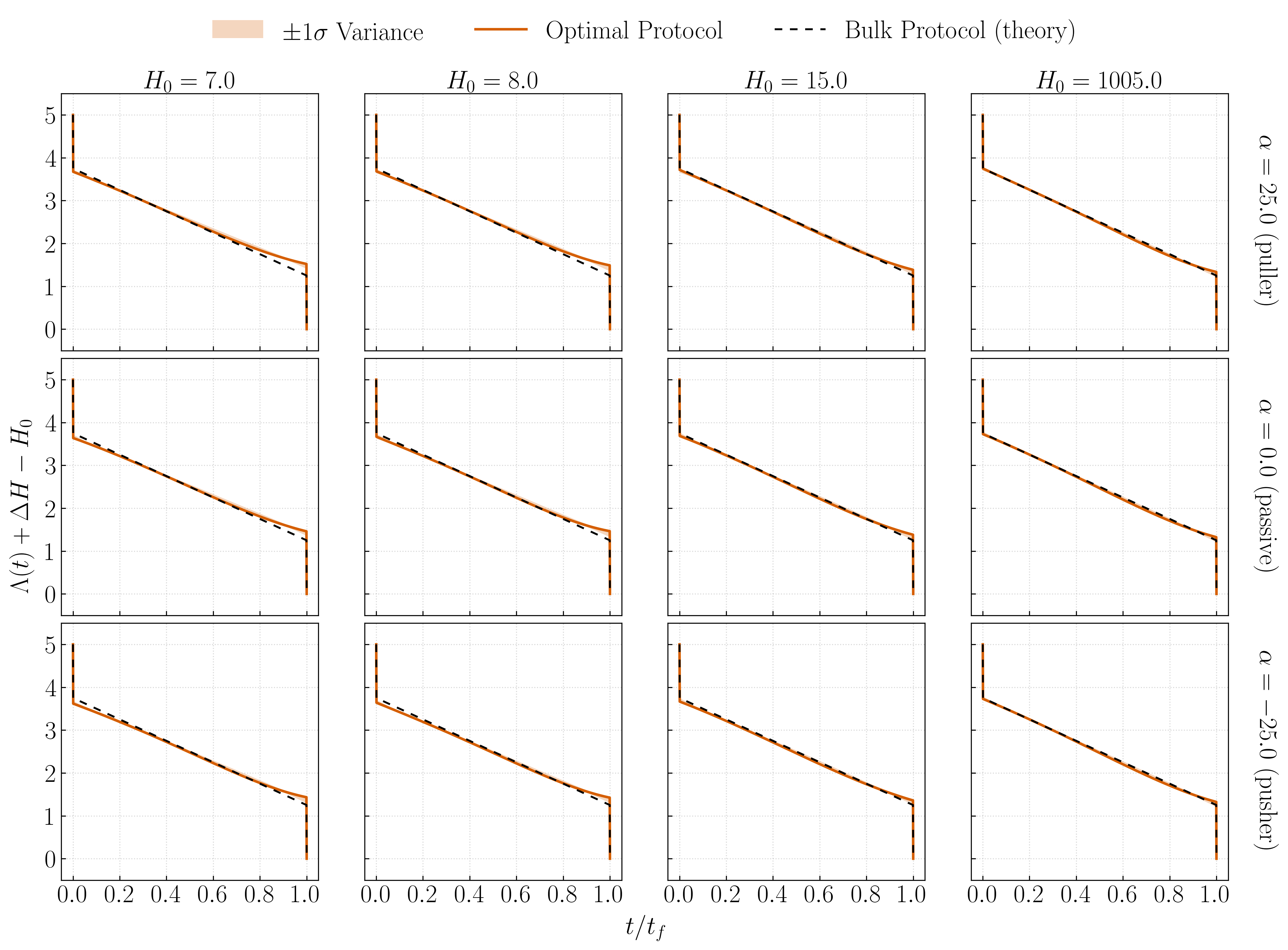}
    \caption{Optimal open-loop protocols for transport \textit{towards} the wall.
    Layout as in Fig.~\ref{fig:protocol_grid}.
    Deviations from the bulk prediction are smaller than for the away direction
    at the same $H_0$ and are concentrated near $t = t_f$.}
    \label{fig:protocol_grid_towards}
\end{figure*}

\begin{figure*}[t]
    \centering
        \includegraphics[width=0.46\textwidth]{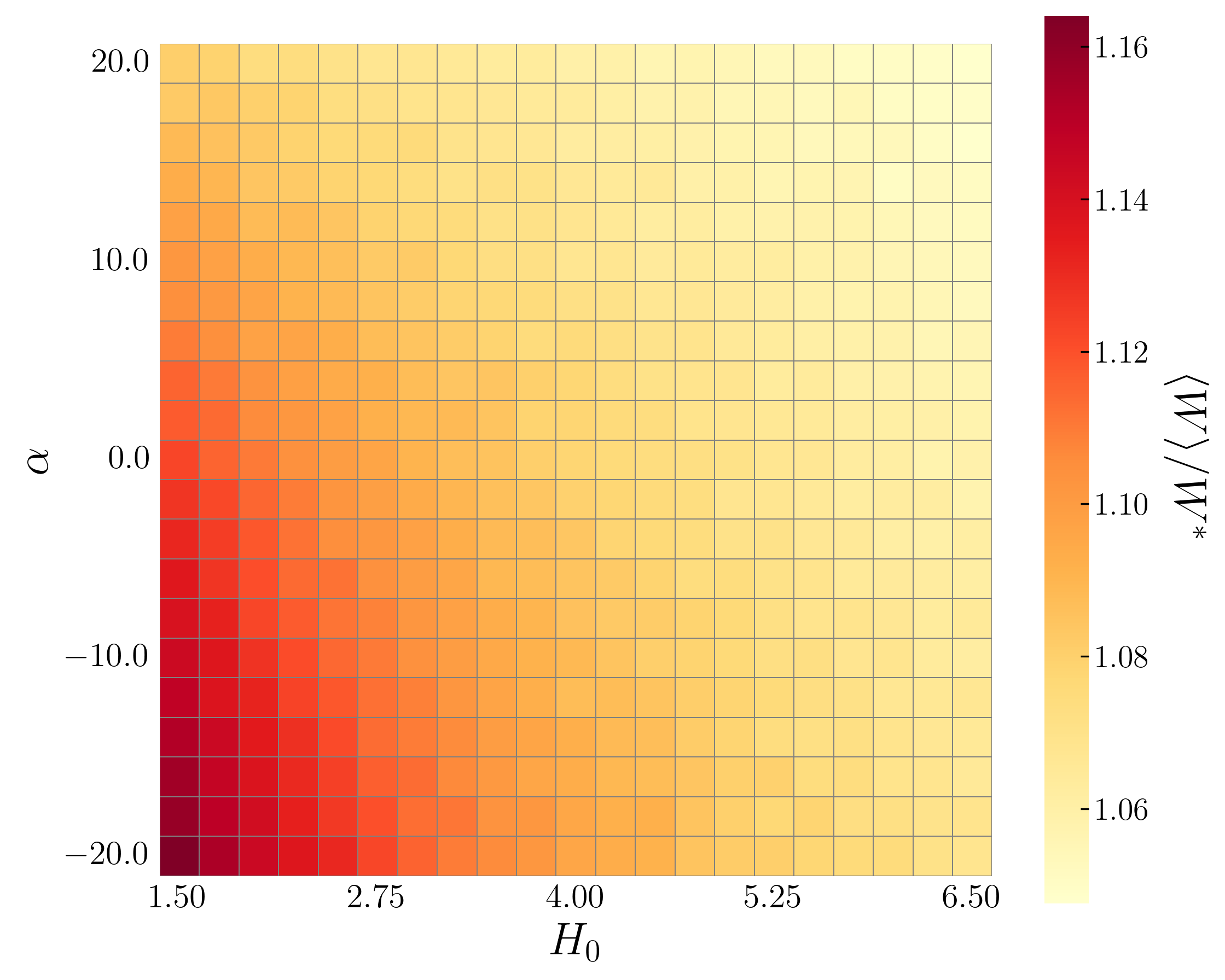}
    \hspace{0.9em}
        \includegraphics[width=0.46\textwidth]{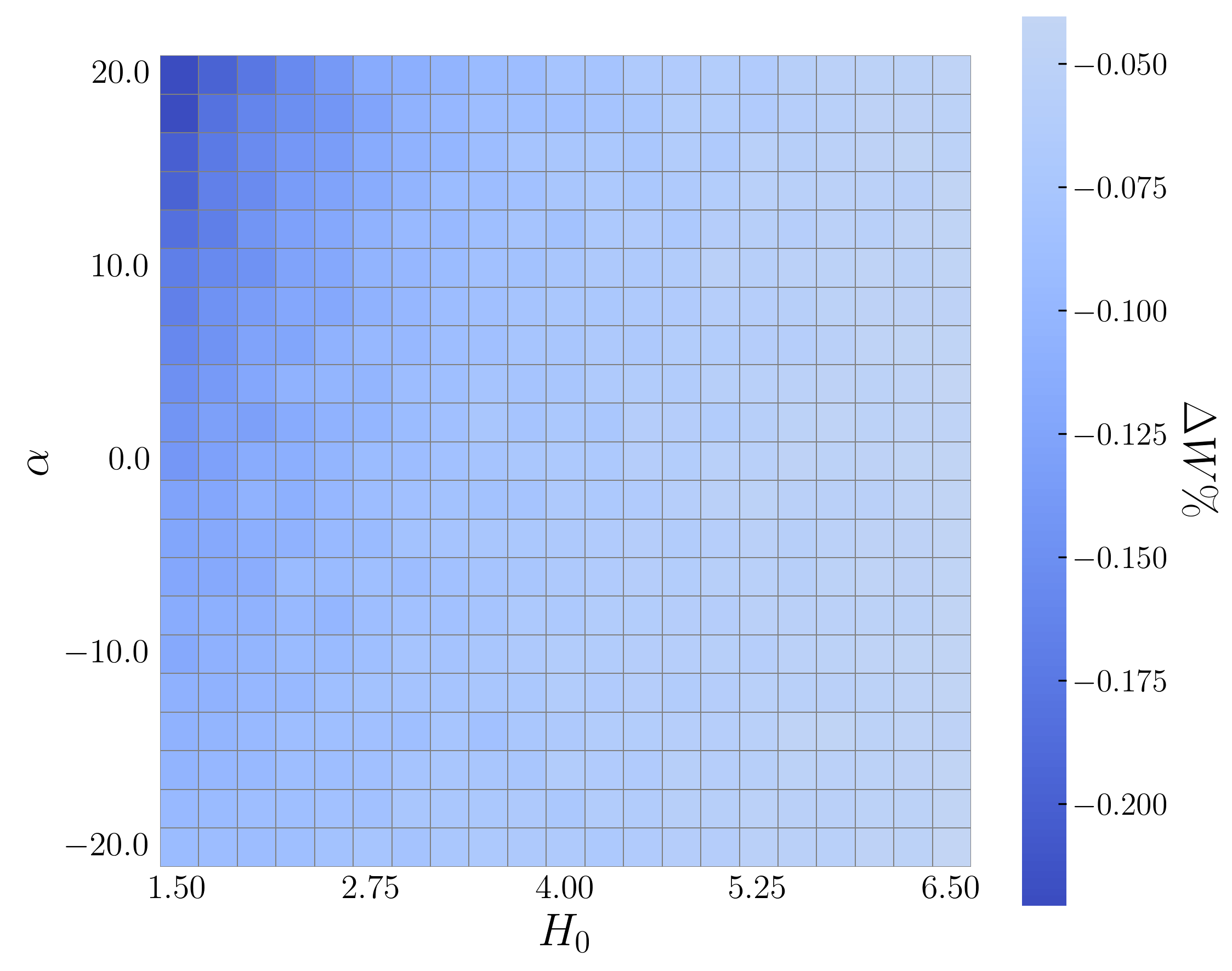}
    \caption{Heatmaps of ADAM optimization results for towards-wall transport.
    LEFT: mean thermodynamic work $\langle W \rangle$; layout as in
    Fig.~\ref{fig:heatmaps_away}a.
    The work has been normalised with the bulk value $W^*$ defined in Eq.\ref{eq:seifert_work}.
    RIGHT: percentage difference $\Delta W\%$; layout and sign convention as in
    Fig.~\ref{fig:heatmaps_away}b.
    reflecting the late-time onset of near-wall effects
    }
    \label{fig:heatmaps_towards}
\end{figure*}

The optimal protocols and work heatmaps for towards-wall transport are presented
here to complement the away-wall results of Section~\ref{sec:results}.
In Fig.~\ref{fig:protocol_grid_towards} we show the optimal trap protocols over the
\textcolor{black}{$4 \times 3$}
 parameter grid, and in Fig.~\ref{fig:heatmaps_towards} we show the
corresponding work heatmaps.
Comparing the two grids, the towards protocols deviate less from the bulk Schmiedl--Seifert
prediction than the away protocols at the same $H_0$, consistent with the
symmetry breaking discussed in Section~\ref{sec:symmetry}: the particle starts far
from the wall where the bulk solution is nearly optimal, and corrections accumulate
only in the final phase as the wall is approached.
The percentage work savings relative to the bulk protocol are correspondingly
smaller across the full parameter space, and the $\alpha$-dependence is reversed: pullers
($\alpha > 0$) are now assisted during towards transport while pushers ($\alpha < 0$) face
additional resistance.

 For the towards journey, the particle starts where the Schmiedl--Seifert protocol is nearly
optimal, and corrections accumulate only in the final phase as the wall is
approached. See Fig.~\ref{fig:protocol_grid_towards} for towards wall transport protocols. 
For activity, the $\alpha$-dependence is reversed for towards transport: pullers
($\alpha > 0$) are now assisted during towards transport while pushers ($\alpha < 0$) face
additional resistance.
This direction-reversal of the activity effect, together with the protocol-shape
asymmetry, completes the characterization of the broken time-reversal symmetry.
The optimized protocols deviate less from the bulk prediction 
at the same $H_0$, and the heatmap shows correspondingly lower absolute
work.

\section{Benchmarks and estimations}
\subsection{The bulk limit}
\label{app:seifert}
In the limit $H_0 \to \infty$, we have ${\mu} \to \mu_0$ and $v^{\mathcal A} \to 0$,
and thus, Eq.~(\ref{eq:mainLE}) reduces to the motion of an overdamped colloidal particle in a harmonic trap in the bulk fluid, which was studied by Schmiedl 
and Seifert~\cite{schmiedl2007optimal}.
By variational calculus, the minimum-work protocol for $0 < t < t_f$ is the linear ramp (along with instantaneous jumps at the endpoints, as described below)
\begin{equation}
    \lambda^*(t) = \lambda_i + (\lambda_f - \lambda_i)\frac{t+1}{t_f+2}
    \label{eq:seifert_protocol}
\end{equation}
\textcolor{black}{We note that in Ref.\cite{schmiedl2007optimal}, $\lambda_i$ was considered to be origin without the loss of generality since the protocol was for transport of the particle in the bulk fluid. Thus, the above expression appears without $\lambda_i$ in Ref.\cite{schmiedl2007optimal}.
In contrast, the translation invariance is lost for motion of colloidal particle near a plane wall. Thus, we need to have an explicit form of $\lambda_i$ in the optimal protocol. }
In addition:  $\lambda^*(0^+) = \lambda_i$ and $\lambda^*(t_f^-) = \lambda_f$, accompanied
by instantaneous jumps at the endpoints whose amplitude is set by the boundary
conditions of the Euler--Lagrange equation.
The corresponding minimum mean work is
\begin{equation}
    W^* = \frac{(\lambda_f - \lambda_i)^2}{t_f + 2},
    \label{eq:seifert_work}
\end{equation}
and, not unexpectedly, $W^* \to 0$ as $t_f \to \infty$.
For the parameters used throughout ($\lambda_f - \lambda_i= 5b$, $t_f = 2\tau$)
this gives $W^* = 6.25$.
The dashed lines in all protocol figures show this solution at the relevant
boundary conditions; the result of method in section \ref{sec:model} at $H_0 = 1000$ is compared against
$W^*$ as the primary validation of the method (Section~\ref{sec:validation}).
 
It is worth noting that the bulk solution possesses a time-reversal symmetry:
the protocol for transport from $H_0$ to $H_0+\Delta H$
is the exact time-reversal of the return protocol, and both achieve the same $W^*$.
This symmetry follows from the spatial uniformity of the bulk dynamics and is
broken due to activity, as shown in Section~\ref{sec:activity}.

%
\subsection{Evaluated quantities}

Two work values are reported for each grid point.
$\langle W \rangle $ is the mean thermodynamic work of the best Adam-evolved
protocol, evaluated retrospectively at $N_\text{traj} = 100{,}000$ trajectories
using the Euler-maruyama integrator (Eq.~\ref{eq:em}).
$W_\text{SS}$ is the mean work of the Schmiedl--Seifert linear-ramp protocol
(Eq.~\ref{eq:seifert_protocol}) evaluated through the same near-wall simulator
at the same $100{,}000$ trajectories and the same random seed.
Since both quantities are computed with identical near-wall physics, their
difference
\begin{equation}
\label{eq:deltaw}
    \Delta W\% =
    \frac{ 
    W_\text{SS} -\langle W \rangle
    }{W_\text{SS} }\times 100\%
\end{equation}
measures the genuine thermodynamic
gain from optimising the protocol shape for the near-wall environment.
In the bulk limit $H_0 = 1000$ both quantities recover $W^* = 6.25$,
confirming that the Schmiedl--Seifert protocol remains optimal when wall effects are absent.

\begin{figure*}[t]
    \centering
    \includegraphics[width=0.96\textwidth]{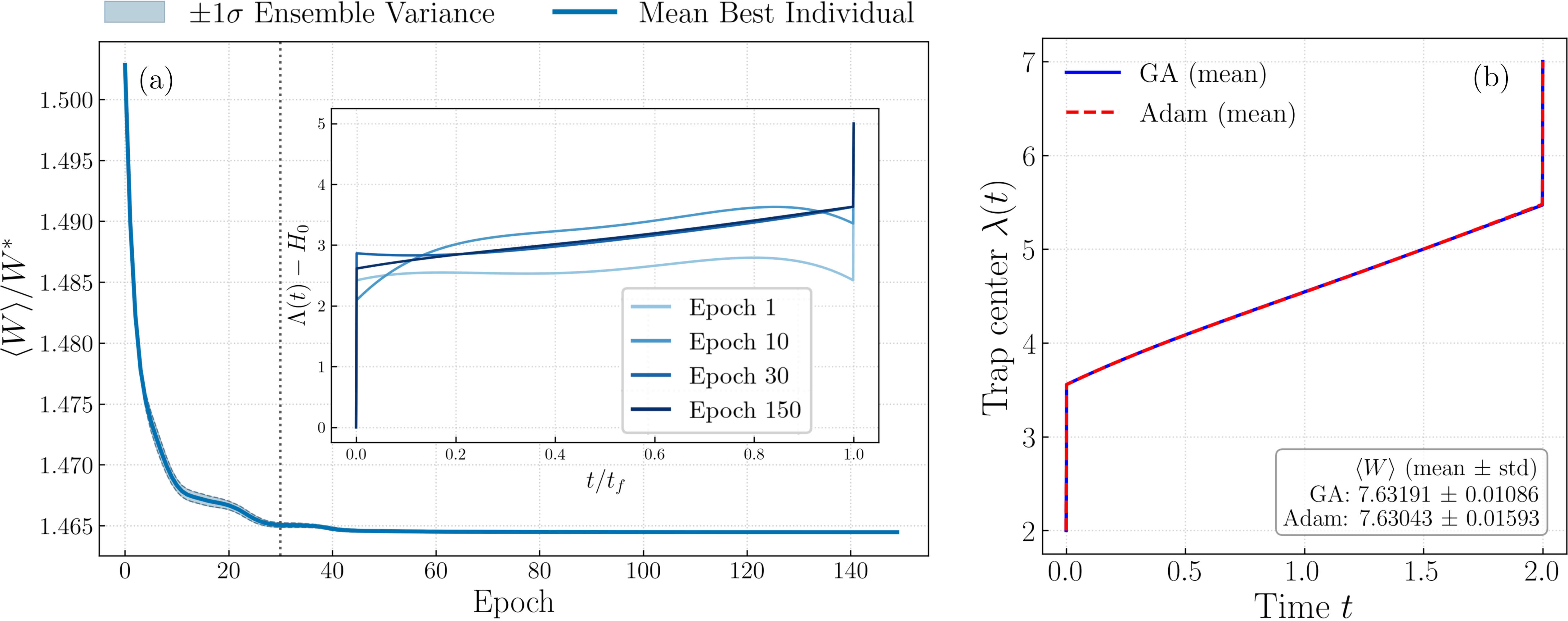}
    \caption{(a) Convergence of Adam over 150 epochs across 30 independent
    trials at $H_0 = 2.0$, $\alpha = 25$.
    Solid line indicates the mean best-individual work $\langle W \rangle$.
Shaded band show ensemble variance ($\pm 1\sigma$ ) 
across trials.
(b) shows that both GA and Adam (gradient-based) method converge to the same solution. 
    }
    \label{fig:hero_convergence}
\end{figure*}

\begin{table}[h]
\caption{Ensemble convergence diagnostics for away-from-wall transport,
computed from 30 independent trials per grid point.
CV: coefficient of variation $100\times\sigma/\mu$ of $\langle W \rangle $.
$L_2$ RMSE ($\times10^{-2}$): mean pairwise protocol distance; $\pm$ values
are the standard deviation across pairs.}
\label{tab:convergence}
\begin{ruledtabular}
\begin{tabular}{lccccc}
 & & \multicolumn{4}{c}{$H_0$} \\
\cline{2-6}
$\alpha$ & Metric & $2.0$ & $3.0$ & $10.0$ & $1000.0$ \\
\hline
\multirow{2}{*}{$+25$}
  & CV        & 0.17 & 0.19 & 0.16 & 0.14 \\
  & $L_2$     & $1.35\pm0.42$ & $1.35\pm0.45$ & $1.50\pm0.63$ & $1.33\pm0.48$ \\
\hline
\multirow{2}{*}{$0$}
  & CV        & 0.19 & 0.22 & 0.17 & 0.18 \\
  & $L_2$     & $1.48\pm0.50$ & $1.47\pm0.52$ & $1.24\pm0.46$ & $1.18\pm0.46$ \\
\hline
\multirow{2}{*}{$-25$}
  & CV        & 0.17 & 0.20 & 0.16 & 0.20 \\
  & $L_2$     & $1.32\pm0.42$ & $1.28\pm0.42$ & $1.17\pm0.45$ & $1.32\pm0.44$ \\
\end{tabular}
\end{ruledtabular}
\end{table}

\subsection{Ensemble Statistics for away-from-wall transport}
\label{sec:ensemeble}
The ensemble statistics across the full \textcolor{black}{$4\times3$} parameter grid are summarized
in Table~\ref{tab:convergence}.
The coefficient of variation $\text{CV} = 100\times\sigma/\mu$, where $\sigma$ and
$\mu$ are the standard deviation and mean of $\langle W \rangle $ across the 30
trials, measures the trial-to-trial variability in the evolved work value.
Values below $0.14\%$ for all grid points confirm that the Adam converges to the same
work value regardless of the random initialization of the population.
The protocol geometric variance, measured by the $L_2$ RMSE between all pairs of
optimal protocols in function space, tests whether this consistency in work value
reflects genuine convergence to a common protocol shape or merely accidental
degeneracy among distinct solutions.
Values of order $2$--$3\times10^{-2}$ across the grid indicate that all 30 trials
converge to functionally similar protocols, ruling out the latter possibility.
Together, the low CV and low $L_2$ RMSE confirm that the Adam landscape has a single
well-defined minimum at each grid point and that the optimization reliably locates
it.


\subsection{Convergence of results}

%
%
\begin{figure*}[t]
    \centering
    \includegraphics[width=0.9\textwidth]{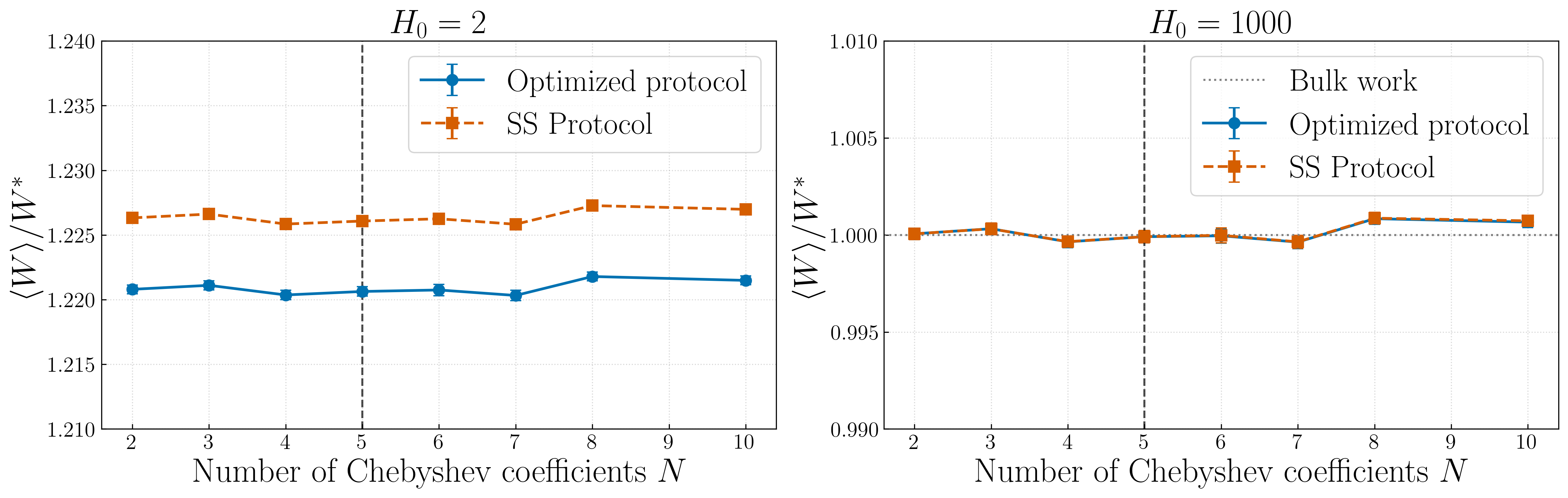}
    \caption{ Convergence of mean work with the number of Chebyshev coefficients, $N$.
    }
    \label{fig:hero_convergence2}
\end{figure*}
\begin{figure*}[t]
    \centering
    \includegraphics[width=0.889\textwidth]{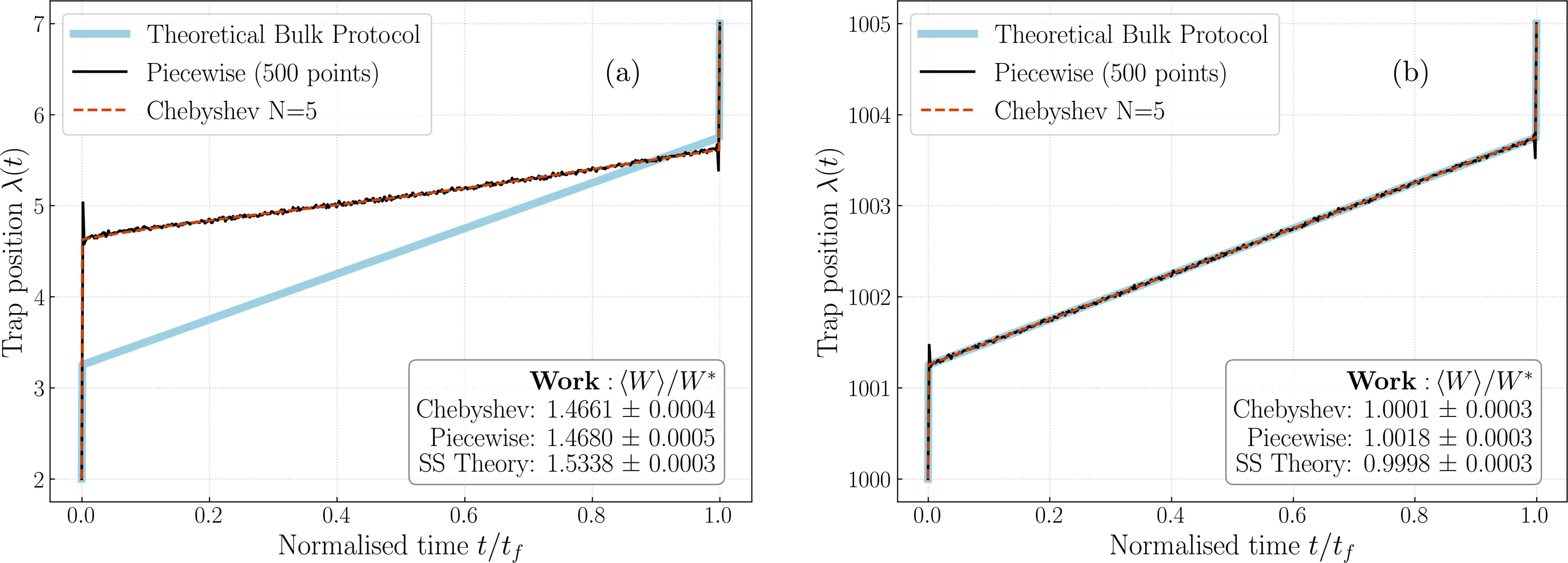}
    \caption{Comparison of the Chebyshev polynomials against a piecewise linear parameterization. Panel (a) is for $H_0=2$ and panel (b) is for $H_0=1000$. In the latter, all the protocols match in the bulk, while there is an extra jump for piecewise linear parameterization at the start and end of the protocol. }
    \label{fig:cheb}
\end{figure*}
Fig.~\ref{fig:hero_convergence} shows the convergence history for the
representative grid point $H_0 = 2$, $\alpha = 25$.
The best protocol from each trial is retrospectively evaluated at
$100{,}000$ trajectories per epoch using a shared fixed noise seed,
a technique known as Common Random Numbers (CRN), which ensures that work
values across trials and epochs are directly comparable.
The algorithm reaches a stable minimum by approximately epoch~50, with
negligible improvement thereafter.

Figure \ref{fig:hero_convergence2} shows the convergence with number of 
Chebyshev coefficients, $N$. It can be seen that the convergence remains robust on increasing $N$. We use $N=5$ in this paper. 

In Fig.\eqref{fig:cheb}, we compare different choices for protocol representation. In particular, we compare Chebyshev polynomials against a piecewise linear parameterization. This is done for two values of $H_0$. We find that the 
Chebyshev polynomial is the optimal choice as it matches the bulk solution exactly.

\end{document}